\documentclass[iop]{emulateapj}
\usepackage{apjfonts}

\newcommand{\myemail}{kwajima@shao.ac.cn}

\shorttitle{SHORT-TERM RADIO VARIABILITY IN 1H~0323+342}
\shortauthors{WAJIMA ET~AL.}

\begin{document}

\title{Short-Term Radio Variability and Parsec-Scale Structure \\
in a Gamma-Ray Narrow-Line Seyfert 1 Galaxy 1H~0323+342}

\author{Kiyoaki Wajima\altaffilmark{1}}

\author{Kenta Fujisawa\altaffilmark{2,3}}

\author{Masaaki Hayashida\altaffilmark{4}}

\author{Naoki Isobe\altaffilmark{5}}

\author{Takafumi Ishida\altaffilmark{3}}

\author{Yoshinori Yonekura\altaffilmark{6}}

\altaffiltext{1}{Shanghai Astronomical Observatory, Chinese Academy of
Sciences, 80 Nandan Road, Xuhui District, Shanghai 200030, China;
\myemail}

\altaffiltext{2}{The Research Institute for Time Studies, Yamaguchi University,
1677-1 Yoshida, Yamaguchi, Yamaguchi 753-8511, Japan}

\altaffiltext{3}{Graduate School of Science and Engineering, Yamaguchi
University, 1677-1 Yoshida, Yamaguchi, Yamaguchi 753-8512, Japan}

\altaffiltext{4}{Institute for Cosmic Ray Research, The University of Tokyo,
5-1-5 Kashiwanoha, Kashiwa, Chiba 277-8582, Japan}

\altaffiltext{5}{The Institute of Space and Astronautical Science, Japan
Aerospace Exploration Agency, 3-1-1 Yoshinodai, Chuo-ku, Sagamihara, Kanagawa
252-5210, Japan}

\altaffiltext{6}{Center for Astronomy, Ibaraki University, 2-1-1 Bunkyo, Mito,
Ibaraki 310-8512, Japan}

\begin{abstract}
We made simultaneous single-dish and very long baseline interferometer (VLBI)
observations of a narrow-line Seyfert 1 galaxy (NLS1) 1H~0323+342, showing
gamma-ray activity revealed by {\it Fermi}/LAT observations.
We found significant variation of the total flux density at 8~GHz on the time
scale of one month by the single-dish monitoring.
The total flux density varied by 5.5\% in 32 days, which is comparable to the
gamma-ray variability time scale, corresponding to the variability brightness
temperature of $7.0 \times 10^{11}$~K.
The source consists of central and southeastern components on the parsec (pc)
scale.
The flux of only the central component decreased in the same way as the total
flux density, indicating that the short-term radio variability, and probably
the gamma-ray emitting region, is associated with this component.
From the VLBI observations we obtained the brightness temperatures of greater
than $(5.2 \pm 0.3) \times 10^{10}$~K, and derived the equipartition Doppler
factor of greater than 1.7, the variability Doppler factor of 2.2, and the
8~GHz radio power of $10^{24.6}$~W~Hz$^{-1}$.
Combining them we conclude that acceleration of radio jets and creation of
high-energy particles are ongoing in the central engine, and that the apparent
very radio-loud feature of the source is due to the Doppler-boosting effect,
resulting in the intrinsic radio loudness to be an order of magnitude smaller
than the observed values.
We also conclude that the pc-scale jet represents recurrent activity from the
the spectral fitting and the estimated kinematic age of pc- and kpc-scale
extended components with different position angle.
\end{abstract}

\keywords{galaxies: active --- galaxies: individual (1H~0323+342) ---
galaxies: Seyfert --- radio continuum: galaxies --- techniques:
interferometric}

\section{Introduction}
\label{sec:Section1}

Gamma-ray emission from active galactic nuclei (AGNs) is one of the important
properties in terms of activities and energetics of the central engine and
emergence of relativistic jets.
In the {\it EGRET} (Energetic Gamma-Ray Experiment Telescope) era, nearly a
hundred of gamma-ray emitting AGNs were known
\citep{Hartman99,Sowards03,Sowards04}.
Those are predominantly radio-loud AGNs and categorized as blazars, suggesting
a close connection between gamma-ray and radio emission.
On the other hand, observations with the {\it Fermi Gamma-ray Space Telescope}
(hereafter {\it Fermi}) have resulted in the identification of more than a
thousand AGNs, some of which are categorized as non-blazar active galaxies or
radio galaxies \citep{Nolan12}.
Moreover, {\it Fermi} has also revealed the existence of new classes of
gamma-ray emitting AGNs, one of which is narrow-line Seyfert 1 galaxy (NLS1).
NLS1 is a subclass of AGNs and identified by their optical properties; narrow
permitted lines, ${\mathrm{FWHM(H \beta)}} < 2000$~km~s$^{-1}$, emitted from
the broad line region, [\ion{O}{3}]$/{\mathrm{H}} \beta < 3$, and a bump due to
\ion{Fe}{2} \citep{Osterbrock85,Pogge00}.

First detection of gamma-ray emission in a radio-loud NLS1 by {\it Fermi} was
made for PMN J0948+0022 \citep{Abdo09a}.
Intensive observations by the Large Area Telescope (LAT) onboard {\it Fermi}
have revealed new detections of gamma-rays of other three radio-loud NLS1s
\citep{Abdo09b}.
Seven NLS1s are listed as gamma-ray sources detected by {\it Fermi}/LAT to date
\citep{Foschini11} and the number of gamma-ray NLS1s is increasing\footnote{See
http://www.brera.inaf.it/utenti/foschini/gNLS1/catalog.html}.
These are surprising discoveries because most of NLS1s are hosted in a spiral
galaxy, which usually does not have relativistic jets, while blazars are hosted
in an elliptical galaxy.
Although milliarcsecond (mas)-scale images have been obtained for several
gamma-ray NLS1s with very long baseline interferometer (VLBI) observations
\citep{Doi06b,Doi07,Giroletti11,Linford12,Orienti12,D'Ammando12,D'Ammando13}
and some of them have revealed presence of a closely aligned relativistic jet,
there are still large uncertainties for the derived parameters in gamma-ray
NLS1s and it is therefore important to investigate the parsec (pc)-scale
properties.

NLS1s are also considered to have high-mass accretion rates, close to the
Eddington limit, and lower black hole masses ($M_{\mathrm{BH}} \sim 10^{5 - 7}
M_{\sun}$) compared to other classes of AGNs \citep{Boroson02}.
Considering the relation among radio loudness, the black hole mass, and the
accretion rate \citep[][and references therein]{Greene06,Zhou07}, those
properties imply that NLS1s are generally radio-quiet objects and have weak or
no jet activity \citep{Maccarone03}.
In fact, previous studies show that the fraction of radio-loud NLS1s is low
(about 7\%) compared to other AGN classes \citep{Komossa06} and NLS1s have
lower jet activities compared to other types of radio-loud AGNs
\citep{Greene06,Zhou06}.
NLS1s are thus considered to be at the high/soft state containing a slim disk
while blazars and flat-spectrum radio quasars are at the low/hard state, in the
analogy of Galactic X-ray binaries \citep{Mineshige00,Maccarone03}.
It is interesting to study relation between the state transition and the jet
activity in AGNs since we can import knowledge of the state transition for
Galactic X-ray binaries to AGNs.
Again, multiepoch and high-resolution radio observations are essential to
investigate them.

In this paper we report the results of simultaneous single-dish and very long
baseline interferometer (VLBI) observations of 1H~0323+342 at 8~GHz, including
the archival data obtained by the Very Long Baseline Array (VLBA).
1H~0323+342 is one of seven gamma-ray emitted NLS1s detected by {\it Fermi}/LAT
\citep{Abdo09b}.
An optical observation by the {\it Hubble Space Telescope} shows that the
source is hosted in a spiral galaxy containing single spiral arm
\citep{Zhou07}, while another results claim that the host galaxy of the source
has a ring structure triggered by an interacting/merging process
\citep{Anton08}.
$M_{\mathrm{BH}}$ is estimated to be a few times $10^7 M_{\sun}$ with a few
different methods using the width and luminosity of the H$\beta$ line and the
continuum luminosity at 5100~\AA \citep{Zhou07}.
This is in good agreement with $M_{\mathrm{BH}} = 10^{7.0}~M_{\sun}$ from the
parameter used to model the spectral energy distribution \citep{Abdo09b}.
\citet{Abdo09b} estimate the accretion disk luminosity to be $1.4 \times
10^{45}$~erg~s$^{-1}$, or $0.9 L_{\mathrm{Edd}}$, where $L_{\mathrm{Edd}}$ is
the Eddington luminosity.
Although these characteristics are typically seen among NLS1s, the source has
extremely high radio loudness $R_{1.4} = f_{\mathrm{1.4 GHz}}/
f_{\mathrm{440 nm}} = 318$ \citep{Foschini11}, or $R_5 = f_{\mathrm{5 GHz}}/
f_{\mathrm{440 nm}} = 246$ \citep{Doi12}, and is listed in one of 29 very
radio-loud ($R > 100$) NLS1s \citep{Foschini11}, implying activities due to
relativistic jets in the inner region.
A redshift, $z$, of $0.0629 \pm 0.0001$ is measured by \citet{Zhou07} and we
adopt this value, although some of previous articles adopt $z = 0.061$ reported
by \citet{Marcha96}.
1H~0323+342 is the nearest object among gamma-ray NLS1s, allowing us to
investigate the inner region of the source with higher angular resolution.

The content of the paper is as follows.
Our observations and data reduction procedures are described in
Section~\ref{sec:Section2}.
We present the results of single-dish and VLBI observations in
Section~\ref{sec:Section3}.
In Section~\ref{sec:Section4}, we describe the results of the archival VLBA
observations.
In Section~\ref{sec:Section5}, we discuss the observed variability feature,
relation to general properties of NLS1s, and possible connections between
gamma-ray emission and the pc-scale structure.
The achievements of our study are summarized in Section~\ref{sec:Section6}.
Throughout this paper, we define the spectral index, $\alpha$, as $S_{\nu}
\propto \nu^{+\alpha}$, where $S_{\nu}$ is the flux density at the frequency
$\nu$, and we adopt a $\Lambda$CDM cosmology with $H_0 =
71$~km~s$^{-1}$~Mpc$^{-1}$, $\Omega_{\Lambda} = 0.73$, and $\Omega_{\mathrm{M}}
= 0.27$ from the results of the {\it Wilkinson Microwave Anisotropy Probe}
\citep{Komatsu09}, corresponding to an angular-to-linear scale conversion of
1.20~pc~mas$^{-1}$ for 1H~0323+342.

\section{Observations and Data Reduction}
\label{sec:Section2}

The total flux density of 1H~0323+342 was monitored by Yamaguchi 32~m radio
telescope \citep[hereafter Y32;][]{Fujisawa02} at 25 epochs from 2010 November
9 (2010.858) to 2011 February 5 (2011.099).
The Y32 observations were made typically every two or three days.
The monitoring was done at 8.38~GHz with the total bandwidth of 400~MHz in the
total power mode.
We employed the `Z-scan' method to remove both the gain and atmospheric
fluctuations, and the pointing offsets.
The detailed procedure of the observation and data reduction of the method is
described by \citet{Kadota12}.
The flux density of 1H~0323+342 was measured by comparing the total received
power with that of a strong, nearby radio source 3C~123
\citep[9.310~Jy;][]{Ott94}.

Observations with the Japanese VLBI Network \citep[JVN;][]{Doi06a} were made
simultaneously with Y32 monitoring at three epochs in November 2010.
The observation dates at each epoch are shown in Table~\ref{tbl:Table1}.
Six radio telescopes, VERA
\citep[the VLBI Exploration of Radio Astrometry;][]{Kobayashi03} 4 telescopes
with the antenna diameter of 20~m each, Hitachi 32~m telescope
\citep{Yonekura13}, and Kashima 34~m telescope, were used for the observations,
while Kashima did not participate in epoch 1.
Observations were made in right-circular polarization at 8.424~GHz with the
total bandwidth of 32~MHz.
Observed sources were 1H~0323+342 as a target, J0310+3814 as a gain calibrator
(separation angle of 4\fdg94 from 1H~0323+342), DA~193 as a flux calibrator,
and 3C~84 as a bandpass calibrator.
The total on-source duration for 1H~0323+342 is 443~minutes in all epochs.
The data were correlated using the Mitaka FX correlator \citep{Shibata98}
with the output preaveraging time of 2 seconds.

The JVN data were reduced using the Astronomical Image Processing System (AIPS)
software \citep{Greisen03} for amplitude and phase calibration, and the Caltech
software Difmap \citep{Shepherd97} for imaging and self-calibration.
We did not employ a standard {\it a~priori} amplitude calibration using the
system noise temperature because it was not measured during the observations at
several stations.
Flux densities of each source were therefore determined as follows.
We determined the flux density of DA~193 by single-dish observations with Y32
to be $4.787 \pm 0.027$~Jy on 2010 November 15 and $4.907 \pm 0.040$~Jy on 2010
November 29.
We assumed the correlated flux density of DA~193 with JVN to be same as that
with single-dish measurements since DA~193 can be considered as a point source
within the JVN baseline lengths.
Flux densities were determined comparing cross-correlation amplitudes for each
source with those for DA~193.
To correct variation of the antenna gain due to change of the antenna
elevation, we applied self-calibration to the gain calibrator J0310+3814.
It can be considered as a point source and was observed once an hour with the
on-source duration of 5 minutes per scan.
A fitted curve to the solutions for the gain calibrator was applied to other
sources.

Through the procedure described above, we determined flux densities of
J0310+3814 at epochs 2 and 3 to be comparable to previous results by VLBI
observations at the same frequency.
On the other hand, the flux density of J0310+3814 at epoch 1, obtained by the
AIPS task GETJY, was determined to be twice (1.913~Jy) than that at epochs 2
(0.985~Jy) and 3 (0.989~Jy).
We finally determined a flux density of J0310+3814 at epoch 1 to be same as
that obtained at epoch 2 for the following reasons.
At epochs 2 and 3, both Kashima 34~m and Hitachi 32~m telescopes participate in
the observations.
Those antennas have system equivalent flux densities (SEFDs) to be
approximately 7 times better than those of VERA 4 antennas at 8~GHz.
We therefore consider the results at epochs 2 and 3 to be more likely than that
at epoch 1.
In addition, we obtain similar flux densities at epochs 2 and 3 as mentioned
above.
We therefore consider that J0310+3814 does not show variability during the JVN
campaign.

We exported the calibrated visibility data to Difmap for imaging.
The amplitude calibration error obtained by self-calibration procedure was
10\%, 4\%, and 4\% for each epoch.
To ensure a better angular resolution, we adopt uniform weighting of the data
with gridding weights scaled by amplitude errors raised to the power of $-1$.
The maximum and minimum projected baseline lengths of the JVN observation were
2,290~km or 64~M$\lambda$ (VERA Mizusawa -- VERA Ishigakijima baseline) and
77~km or 2.2~M$\lambda$ (Hitachi -- Kashima baseline), respectively.

\section{Results}
\label{sec:Section3}

\subsection{Single-Dish Monitoring with Yamaguchi 32~m \\ Radio Telescope}
\label{subsec:Section3-1}

\begin{figure}
\epsscale{1.00}
\plotone{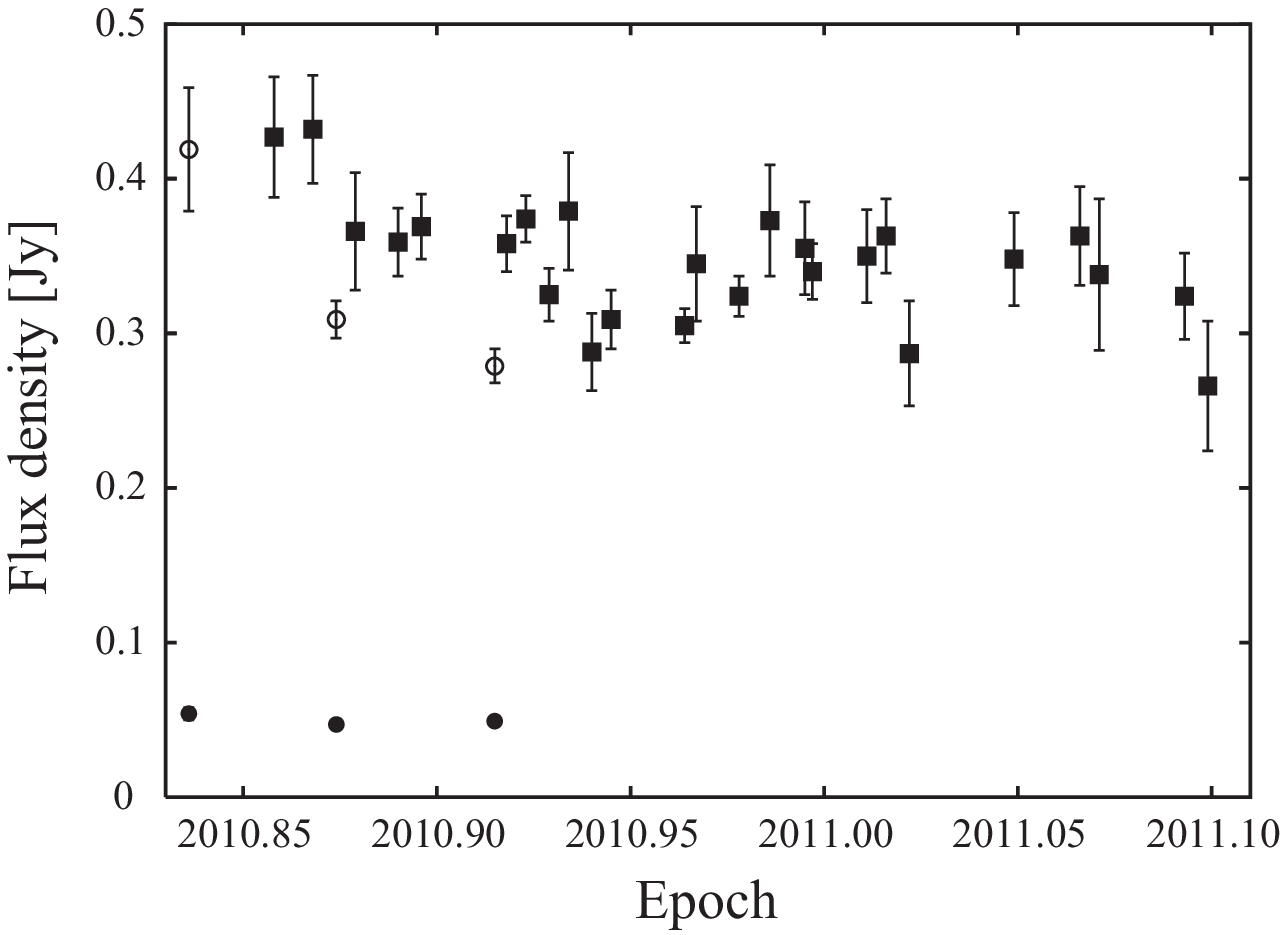}
\caption{8.4~GHz light curve of 1H~0323+342.
The filled squares show the total flux obtained with Yamaguchi 32~m radio
telescope.
Numerical data of the measurements are shown in Table~\ref{tbl:Table2}.
The open and filled circles indicate the flux of the components C and D1,
respectively, obtained with the JVN observations (see also
Section~\ref{subsec:Section3-2} and Table~\ref{tbl:Table1}).}
\label{fig:Figure1}
\end{figure}

Figure~\ref{fig:Figure1} shows the total flux density measurements of
1H~0323+342 by Y32.
Numerical data of the measurements are shown in Table~\ref{tbl:Table2}.
Error bars at each point in Figure~\ref{fig:Figure1} indicate a 1$\sigma$
standard deviation divided by square root of the number of independent
measurements.
The source has also been monitored between 2010.6 to 2011.2 at nine frequencies
from 2.64 to 142~GHz by the F-GAMMA Program \citep{Fuhrmann11,Angelakis12}.
Our results are in good agreement with theirs at 8.35~GHz within the margin of
error.
F-GAMMA results show that the flux density varies almost simultaneously at
observing frequencies from 2.64~GHz to 32.0~GHz \citep{Fuhrmann11}, therefore
we believe that the flux variation observed with Y32 is intrinsic.
The total flux varies significantly during the period of Y32 monitoring
compared with the constant flux.
The maximum and minimum flux densities during Y32 monitoring can be seen as
$432 \pm 24$~mJy on 2010 November 13 and $266 \pm 30$~mJy on 2011 February 5,
respectively, corresponding to a flux decrease of 38\% in 84 days, while the
total flux does not decrease monotonically.
We apply a third-order polynomial to all measurements of Y32 to estimate the
local minimum and maximum of the flux densities (reduced $\chi^2$ = 3.61 with
d.o.f.\ of 23, and 2.33 with d.o.f.\ of 21 for linear and cubic function,
respectively, which is significant at $> 99.9$\% confidence level).
The local minimum of 325~mJy at the epoch 2010.956 and the local maximum of
344~mJy at the epoch 2011.043 are derived from the best-fit curve,
corresponding to the flux variation of 5.5\% in 32 days.
While \citet{Fuhrmann11} pointed out that the source indicates variability on
time scales of months to years, our results clearly show the existence of the
short-term radio variability on the time scale of one month.
Moreover, this is comparable to the (e-folding) gamma-ray variability time
scale of $17.7 \pm 14.4$ days with the two-year monitoring results from 2008
September to 2010 September by {\it Fermi}/LAT \citep{Calderone11}, suggesting
that the source of short-term radio variability is probably associated with
the gamma-ray emitting region.

\subsection{JVN Observation Results}
\label{subsec:Section3-2}

\begin{figure}
\epsscale{0.99}
\plotone{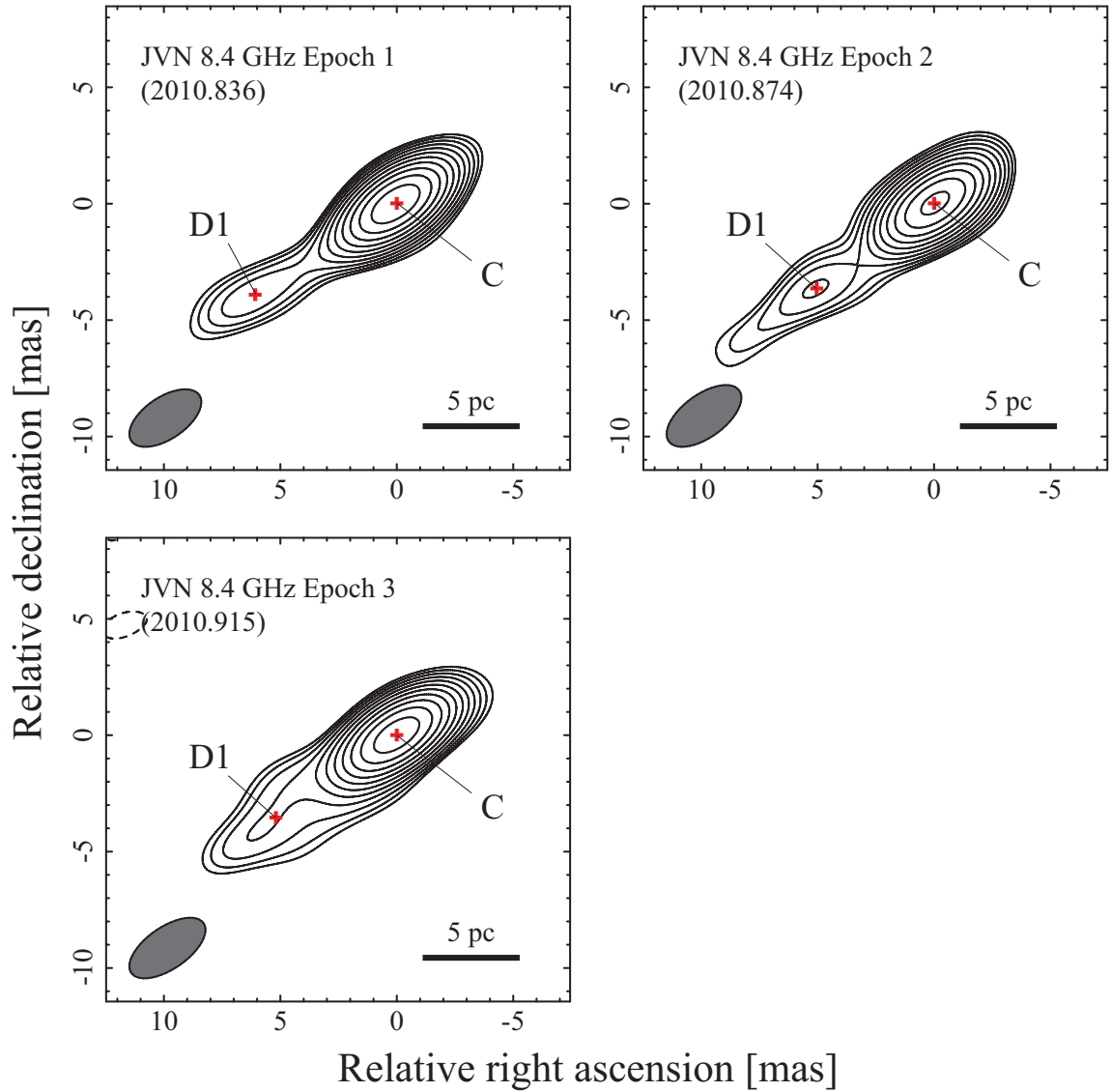}
\caption{VLBI images of 1H~0323+342 at epochs 1 (top left), 2
(top right), and 3 (bottom) obtained by the JVN observations at 8.4~GHz.
The lowest contour is 3 times the off-source rms noise ($\sigma$).
The contour levels are $-3\sigma$, $3\sigma$ $\times$ $(\sqrt2)^n$
($n$ = 0, 1, 2, $\cdot\cdot\cdot$, 10).
Dashed and solid curves show negative and positive contours, respectively.
The restoring beam is indicated at the lower left corner of each image.
The labels C and D1 show the Gaussian model fitting components and the position
of each component is indicated by the cross.
The image descriptions are shown in Table~\ref{tbl:Table1}.}
\label{fig:Figure2}
\end{figure}

Figure~\ref{fig:Figure2} shows images of 1H~0323+342 at each epoch.
The image parameters are shown in Table~\ref{tbl:Table1} in addition to the
total flux density of all CLEAN components.
JVN images could not resolve the central one, containing a few components in
the images obtained by VLBA, as shown in Section~\ref{sec:Section4}.
To quantify the relative location and the flux density of each component of the
JVN images, we modeled the calibrated images with elliptical Gaussian
components.
Table~\ref{tbl:Table3} shows the model fitting results obtained by the AIPS
task JMFIT.
The formal errors of each parameter are estimated using the fomulae from
\citet{Fomalont99}.
The data at all epochs are modeled satisfactorily by two distinct components,
brighter and unresolved component labeled C and weaker component labeled D1,
situated to the southeast of C.
The angular size of C shown in Table~\ref{tbl:Table3} is therefore the upper
limit.
The sum of the flux densities of the components C and D1 is in good agreement
with both the total CLEANed flux and single-dish measurements by Y32, as shown
in Tables~\ref{tbl:Table1} and \ref{tbl:Table2}, while the Very Large Array
(VLA) observations at 1.4~GHz on 2006 February 7 and 2007 January 15 (VLA
observation code; AP501) show the source having a large-scale structure with
the total size of about a hundred kpc \citep{Anton08} and with the flux density
for the extended component of 198~mJy \citep{Doi12}.
We consider that the extended structure has an optically thin spectrum and
therefore shows very weak emission at around 8~GHz.
It should also be noted that the flux density of the component C gradually
decreases similar to the total flux density, while the flux density of the
component D1 seems to be stable, as shown in Figure~\ref{fig:Figure1}.
The results indicate that the short-term radio variability observed with Y32
(and probably the gamma-ray emitting region) is mainly associated with the
central component.

\section{VLBA Archive Data}
\label{sec:Section4}

\begin{figure}
\epsscale{0.99}
\plotone{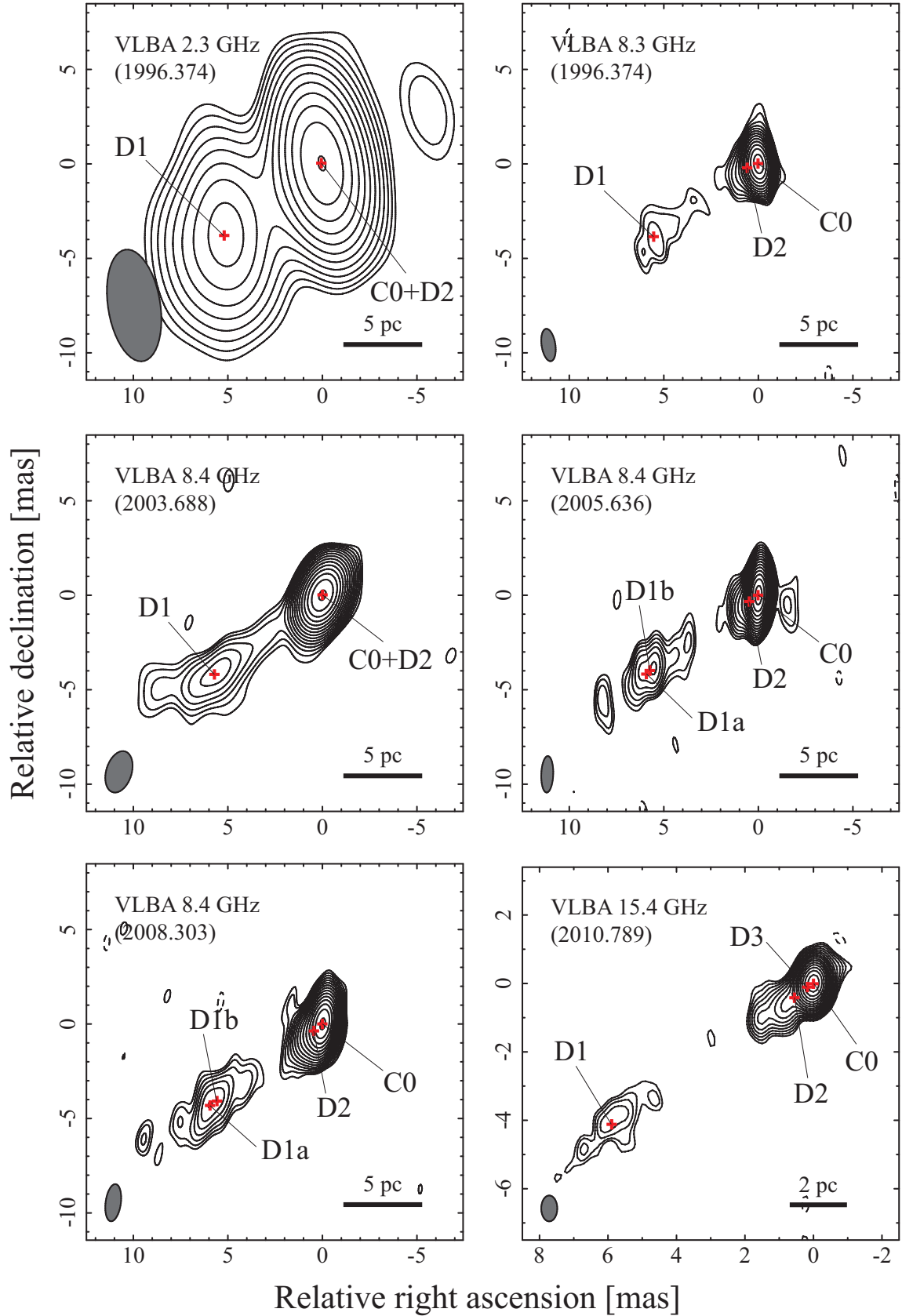}
\caption{VLBI images of 1H~0323+342 obtained by the VLBA archive data.
Top left and bottom right panels show images at 2.3~GHz and 15.4~GHz,
respectively, while others are at 8~GHz.
The lowest contour is 3 times the off-source rms noise ($\sigma$).
Dashed and solid curves show negative and positive contours, respectively.
The restoring beam is indicated at the lower left corner of each image.
The labels C0, D3, D2, and D1(a, b) show the Gaussian model fitting components
and the position of each component is indicated by the cross.
The image descriptions are shown in Table~\ref{tbl:Table4}.}
\label{fig:Figure3}
\end{figure}

To investigate time variation of the pc-scale structure of the source, we
reduced data of the archival VLBA observations at 2, 8, and 15~GHz.
The observations are summarized in Table~\ref{tbl:Table4}.
The simultaneous VLBA observation at 2.3 and 8.3~GHz on 1996 May 16 has been
published by \citet{Beasley02} (VLBA observation code; BB023).
The 15~GHz observation has been carried out as part of the VLBA MOJAVE
Program\footnote{See also http://www.physics.purdue.edu/MOJAVE/}
\citep{Lister09} on 2010 October 15, just half a month before the JVN epoch 1
observation.
Figure~\ref{fig:Figure3} shows images of 1H~0323+342 with VLBA.
Descriptions of each image are shown in Table~\ref{tbl:Table4}.
The size of restoring beam for the BK077 observation is slightly larger than
those for other observations at 8~GHz since the visibilities with Mauna Kea
station, which gives longer baseline data, are excluded because of inadequate
accuracy of the amplitude calibration.
All images shown in Figure~\ref{fig:Figure3} have similar pc-scale structure to
those of the JVN observations, consisting of two distinct components, while
each of them are slightly resolved. 
We modeled the calibrated VLBA images with elliptical Gaussian components.
Table~\ref{tbl:Table5} shows the model fitting results.
The model-fitting procedure and estimation of errors for each parameter are
made in the same manner with those for the JVN data.
We could reconstruct VLBA images with three elliptical Gaussian components C0,
D2, and D1 at 8~GHz, while the component D1 could be resolved into two
components D1b and D1a in the epochs 2005.636 and 2008.303.
In the epochs 1996.374 at 2.3~GHz and 2003.688 the images could be modeled by
two components because of lower angular resolution than other epochs.
The component C in the JVN images was resolved into two components C0 and D2 in
the VLBA images at 8~GHz.
For the 15~GHz image we could find an additional component D3 in the vicinity
of C0.

We also compared the image of JVN epoch 1 with that of the 15~GHz VLBA.
We restored the synthesized beam of the VLBA 15~GHz image to the same size with
the JVN epoch 1 observation and obtained the flux density of the central
component as 276~mJy.
The spectral index of $\alpha_{8}^{15} = -0.70$ is thus derived from the
observations by JVN epoch 1 and the VLBA MOJAVE.
The radio power at a rest frequency of 8.0~GHz is estimated to be
$P_{\mathrm{8GHz}} = 10^{24.6}$~W~Hz$^{-1}$, in which a $k$-correction is
applied using $\alpha_{8}^{15}$.

\section{Discussion}
\label{sec:Section5}

\subsection{Brightness Temperature}
\label{subsec:Section5-1}

We have two kinds of radio observations, a single-dish monitoring and VLBI
observations, therefore we can derive brightness temperatures,
$T_{\mathrm{B}}$, using different ways from these results.
$T_{\mathrm{B}}$ in the source's rest frame can be obtained with an image as
\begin{equation}
T_{\mathrm{B,rest}}^{\mathrm{(image)}} = 1.77 \times 10^9 (1+z) \frac{S_{\nu}}
{\nu^2 \theta_{\mathrm{maj}} \theta_{\mathrm{min}}}
\hspace{5mm} {\mathrm{[K]}},
\label{eqn:Equation1}
\end{equation}
where $\theta_{\mathrm{maj}}$~[mas] and $\theta_{\mathrm{min}}$~[mas] are the
FWHM sizes of the Gaussian component in the major and minor axes, respectively,
and $S_{\nu}$~[mJy] is the flux density at an observing frequency $\nu$~[GHz].
Given the model fitting result of $S_{\nu} = 419$~mJy, $\theta_{\mathrm{maj}}
< 1.12$~mas, and $\theta_{\mathrm{min}} < 0.12$~mas for the component C at
epoch 1 which leads the highest $T_{\mathrm{B}}$ among our JVN observations, we
obtain
$T_{\mathrm{B,rest}}^{\mathrm{(image)}} > (8.3 \pm 1.3) \times 10^{10}$~K.
$T_{\mathrm{B,rest}}^{\mathrm{(image)}}$ at the other epochs are shown in
Table~\ref{tbl:Table3}.
\citet{Linford12} made a VLBA observation of the source on 2010 June 30 at
4.8~GHz and estimated the brightness temperature of the core component to be
$2.15 \times 10^{10}$~K, which is comparable to our results.

We can also obtain $T_{\mathrm{B}}$ with the flux variation as
\begin{equation}
T_{\mathrm{B,rest}}^{\mathrm{(var)}} = 4.1 \times 10^{10} \left[
\frac{D_{\mathrm{L}}}{\Delta t (1+z)} \right]^2 \frac{\Delta S_{\nu}}{\nu^2}
\hspace{5mm} {\mathrm{[K]}}
\label{eqn:Equation2}
\end{equation}
\citep{Wagner95}, where $D_{\mathrm{L}}$~[Mpc] is the luminosity distance to
the source, $\Delta S_{\nu}$~[mJy] is a change in the observed flux density at
an observing frequency $\nu$~[GHz] during a period of $\Delta t$~[days].
Given the single-dish monitoring result of $\Delta S_{\nu} = 19$~mJy,
$\Delta t = 32$~days, and applying $D_{\mathrm{L}} = 270$~Mpc, we obtain
$T_{\mathrm{B,rest}}^{\mathrm{(var)}} = 7.0 \times 10^{11}$~K.
Alternatively $T_{\mathrm{B,rest}}^{\mathrm{(var)}}$ can be estimated by
applying the maximum and minimum flux densities during Y32 monitoring,
$\Delta S_{\nu} = 166$~mJy, $\Delta t = 84$~days, to be
$T_{\mathrm{B,rest}}^{\mathrm{(var)}} = 8.9 \times 10^{11}$~K, which is in good
agreement with the above result.

The brightness temperature has been measured for several gamma-ray NLS1s with
radio observations.
\citet{Zhou03} analyzed the VLA observation data of a quasar PMN~J0948+0022
known to have the highest radio loudness in NLS1s, and found the long-term flux
variation with the time scale of a few years.
The brightness temperature was estimated to be $\sim 10^{13}$~K from the
variability.
This high $T_{\mathrm{B}}$ was confirmed by \citet{Doi06b} with their VLBA
images and flux variation at 1.7 -- 15.4~GHz, which result in
$T_{\mathrm{B,rest}}^{\mathrm{(image)}} > 5.5 \times 10^{11} \delta^{-1}$~K
and $T_{\mathrm{B,rest}}^{\mathrm{(var)}} > 3.3 \times 10^{13} \delta^{-3}$~K,
where $\delta$ is the Doppler factor.
$T_{\mathrm{B}}$ of PMN~J0948+0022 was also measured with the global e-VLBI
observation at 22~GHz to be $T_{\mathrm{B,rest}}^{\mathrm{(image)}} \sim
3.4 \times 10^{11}$~K \citep{Giroletti11}.
\citet{Yuan08} compiled previous radio observations for a sample of radio-loud
NLS1s and estimated $T_{\mathrm{B,rest}}^{\mathrm{(var)}}$ for three gamma-ray
sources; $1.1 \times 10^{13}$~K for SBS~0846+513, $5.1 \times 10^{12}$~K for
PMN~J0948+0022, and $3.6 \times 10^{12}$~K for PKS~1502+036.
\citet{D'Ammando13} found significant flux variation by both single-dish flux
monitoring with the Owens Valley Radio Observatory 40~m telescope at 15~GHz and
multiepoch VLA observations at eight frequencies from 1.4 to 22.2~GHz, and
estimated $T_{\mathrm{B,rest}}^{\mathrm{(var)}}$ to be $2.5 \times 10^{13}$~K
from single-dish measurements.
Although our $T_{\mathrm{B}}$ is a few orders of magnitude lower than those
mentioned above, it still exceeds the upper limit of $T_{\mathrm{B}}$ assuming
the condition of energy equipartition, as shown in
Section~\ref{subsec:Section5-2}, suggesting that acceleration of the radio
jet is ongoing in the central engine of the source.

The 8~GHz radio power is estimated to be $P_{\mathrm{8GHz}} =
10^{24.6}$~W~Hz$^{-1}$, as shown in Section~\ref{sec:Section4}, which is much
higher than the radio powers at 1.5~GHz of up to $10^{23.7}$~W~Hz$^{-1}$
measured in a sample of the most radio-luminous starburst galaxies
\citep{Lonsdale93,Smith98a,Smith98b}.
This also supports that not a starburst activity but a nonthermal process in
the central region is responsible for the radio emission from 1H~0323+342.

\subsection{Doppler Factor}
\label{subsec:Section5-2}

If we assume the condition of energy equipartition between particles and
magnetic fields, $T_{\mathrm{B,limit}} \sim 5 \times 10^{10}$~K is derived as a
value for the upper limit in the source's rest frame \citep{Readhead94}.
Observed $T_{\mathrm{B,rest}}^{\mathrm{(image)}}$ for JVN epoch 1 significantly
exceeds the limit.
Thus the equipartition Doppler factor, $\delta_{\mathrm{eq}} =
T_{\mathrm{B,rest}}^{\mathrm{(image)}}/T_{\mathrm{B,limit}}$, of greater than
1.7 should be required to reconcile with our $T_{\mathrm{B}}$.
We can compute the variability Doppler factor as $\delta_{\mathrm{var}} =
(1 + z) (T_{\mathrm{B,rest}}^{\mathrm{(var)}} /
T_{\mathrm{B, limit}})^{1/(3 - \alpha)}$.
Adopting $T_{\mathrm{B,rest}}^{\mathrm{(var)}} = 7.0 \times 10^{11}$~K and
$\alpha_8^{15} = -0.70$, $\delta_{\mathrm{var}} = 2.2$ can be obtained.
Both $\delta_{\mathrm{eq}}$ and $\delta_{\mathrm{var}}$ are in good agreement
with those obtained by the results with the F-GAMMA Program
\citep{Angelakis12}.
These high $\delta$ indicate the existence of highly or mildly relativistic
jet(s) in the inner region of the source.
This is the third source in which the Doppler-beaming effect has been detected
in gamma-ray NLS1 by both direct imaging with VLBI and the flux variation,
following PMN~J0948+0022 \citep{Doi06b} and PKS~1502+036 \citep{D'Ammando13}.

\subsection{Spectral Index Distribution}
\label{subsec:Section5-3}

\begin{figure}
\epsscale{1.00}
\plotone{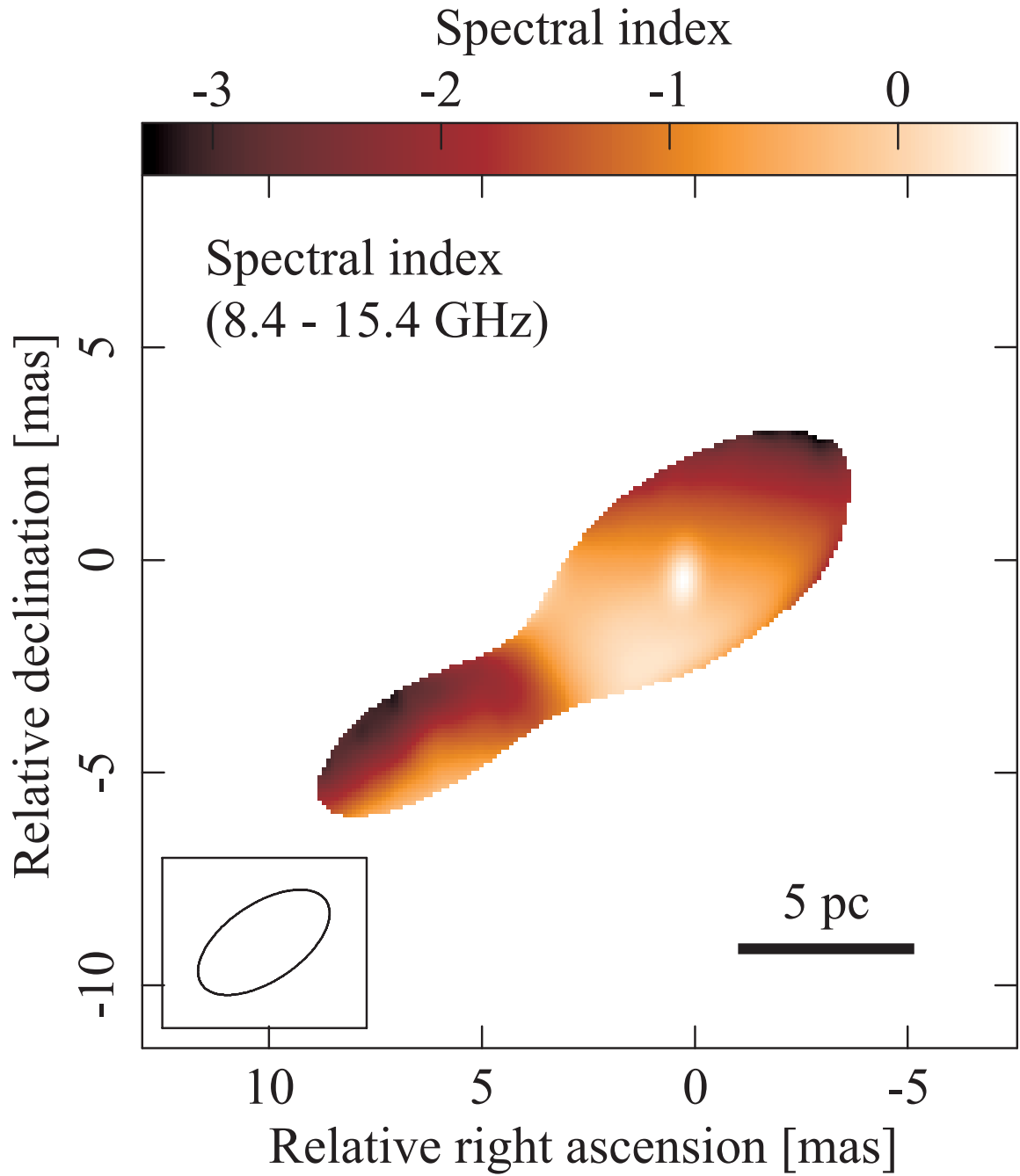}
\caption{Spectral index map of 1H~0323+342 derived from the flux densities at
8.4~GHz (JVN epoch1 on 2010 November 1) and 15.4~GHz (VLBA MOJAVE on 2010
October 15).
The map corresponds to the area greater than 3$\sigma$ noise level in JVN
epoch 1.
The 15.4~GHz map is restored with the same beam size as the 8.4~GHz map, which
is represented in the lower left corner.}
\label{fig:Figure4}
\end{figure}

\begin{figure}
\epsscale{1.00}
\plotone{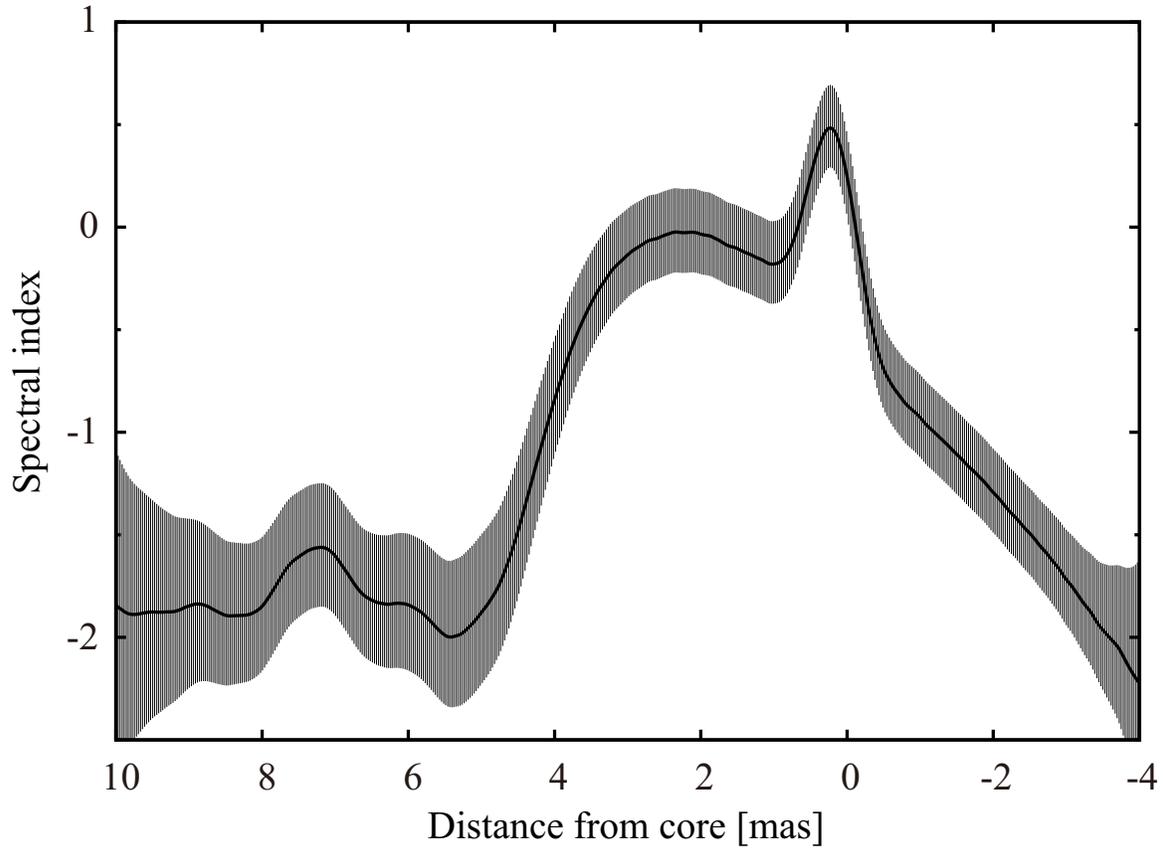}
\caption{Spectral index distribution given by the position angle of 122\fdg8.
The horizontal axis shows the angular separation from the center position of
the component C.
A positive value corresponds to the southeastern direction.}
\label{fig:Figure5}
\end{figure}

Figure~\ref{fig:Figure4} shows the spectral index map of 1H~0323+342 between
JVN epoch 1 at 8.4~GHz (2010 November 1) and the VLBA MOJAVE at 15.4~GHz (2010
October 15).
Both observations do not employ the phase-referencing technique, resulting in
the loss of the absolute position through the self-calibration procedure
\citep{Pearson84,Thompson01}.
We therefore superposed two images with reference to the optically-thin
component D1 since the optically-thick central region generally shows the
frequency-dependent position shift \citep[e.g.,][]{Lobanov98a,Lobanov98b}.
Model fitting to the restored 15.4~GHz MOJAVE image results in the position
offset of the component D1 at 8.4 and 15.4~GHz with respect to the component C
to be $\Delta \alpha = 0.2$~mas and $\Delta \delta = 0.3$~mas in right
ascension and declination, respectively.
The compact central region with a size of 0.6~mas or 0.7~pc shows an optically
thick spectral feature while it becomes steeper along the stream of the jet.
Figure~\ref{fig:Figure5} shows the spectral index distribution given by the
position angle of 122\fdg8, corresponding to the direction of the component D1
with respect to C at JVN epoch 1.
Error bars at each point in Figure~\ref{fig:Figure5} are estimated by the image
rms and the amplitude calibration error of each image.
Maximum of the spectral index at the innermost region is $\alpha = 0.42 \pm
0.20$, showing the central region to have an inverted or flat spectrum at even
higher frequencies.
This feature is similar to that of blazars, suggesting that the source has
activities due to relativistic jets.

\subsection{Proper Motion of Each Component}
\label{subsec:Section5-4}

\begin{figure}
\epsscale{1.00}
\plotone{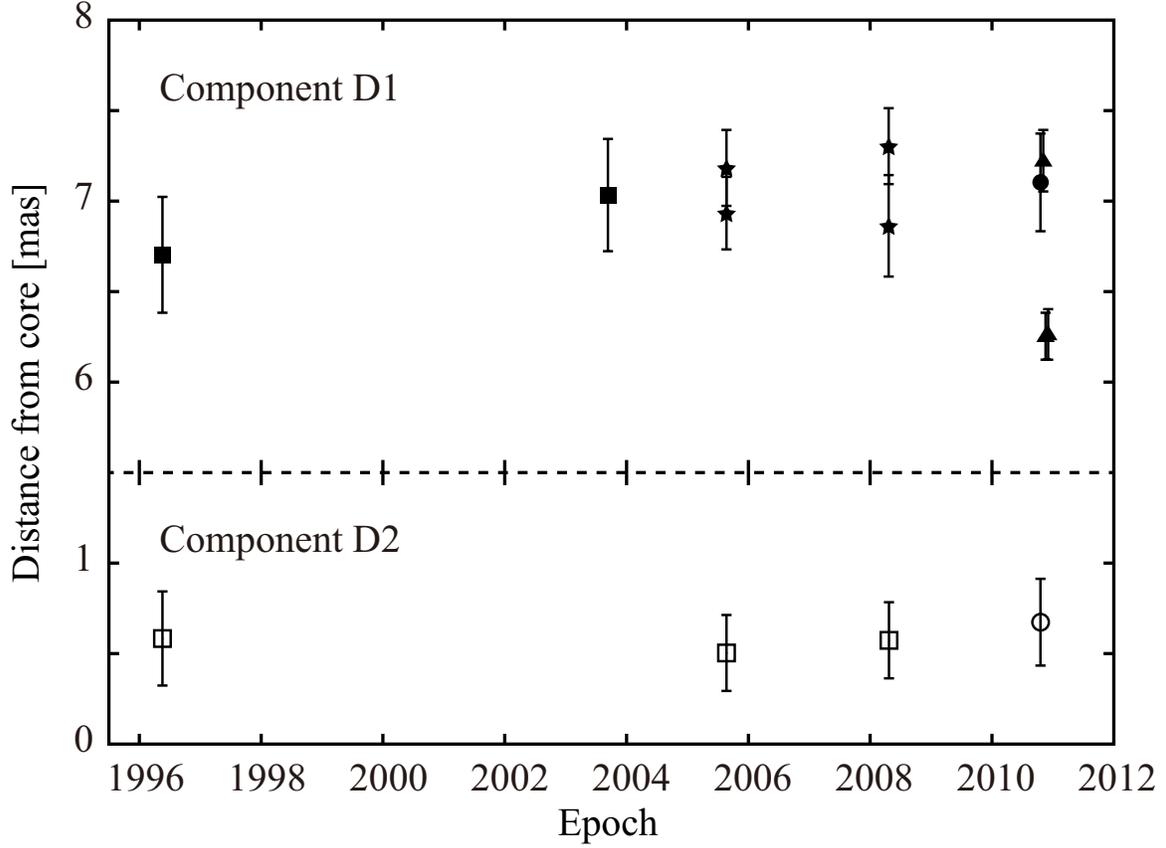}
\caption{Distance of the components D2 and D1(a, b) from the component C (for
JVN observations) or C0 (for VLBA observations) as a function of the observed
epoch.
See Tables~\ref{tbl:Table3} and \ref{tbl:Table5} for numerical values.
The open and filled symbols are for the components D2 and D1, respectively.
Squares and circles represent the VLBA results at 8 and 15~GHz, respectively,
while stars correspond to the components D1a and D1b at epoch 2005.636 and
2008.303.
Triangles represent the JVN results.}
\label{fig:Figure6}
\end{figure}

Figure~\ref{fig:Figure6} shows the distance of the components D2 and D1(a, b)
from the core (C or C0) as a function of the observed epoch.
Although we cannot identify the component D1 with either D1a or D1b, if we
adopt the identification of D1 with D1a the apparent proper motion of $-0.115
\pm 0.083$~mas~yr$^{-1}$ is obtained by a weighted least-squares linear fit to
the component D1, corresponding to the apparent velocity, $v_{\mathrm{app}}$,
of $(-0.45 \pm 0.32)c$.
If we exclude the points for JVN epochs 2 and 3 for a linear fit to D1, the
apparent proper motion of $0.031 \pm 0.011$~mas~yr$^{-1}$ is obtained,
corresponding to $v_{\mathrm{app}} = (0.12 \pm 0.04)c$.
For the component D2, the apparent proper motion is $0.004 \pm
0.008$~mas~yr$^{-1}$, corresponding to $v_{\mathrm{app}} = (0.02 \pm 0.03)c$.
If we omit the data point at 1996.374, the apparent proper motion becomes
$0.033 \pm 0.004$~mas~yr$^{-1}$, corresponding to $v_{\mathrm{app}} =
(0.13 \pm 0.02)c$, which is similar to that for the component D1.
We cannot determine whether the component D1 at the epoch 1996.374 and at
other epochs is identical since no VLBI observation was performed from 1996.374
to 2003.688.
Future multiepoch and intensive VLBI observations will make an identification
and a measurement of precise proper motion of D1.
In either case, these show that both the components D2 and D1 are stationary,
or have very low speeds compared to typical (gamma-ray) blazars
\citep[e.g.,][]{Jorstad01}.

\subsection{Relation to General Properties of NLS1s}
\label{subsec:Section5-5}

As shown in Section~\ref{sec:Section1}, NLS1s are generally considered to have
higher accretion rates and lower jet activities compared to other types of
radio-loud AGNs \citep{Greene06,Zhou06}.
In the analogy of X-ray binaries with $M_{\mathrm{BH}} \sim 10M_{\sun}$, radio
emission is quenched when the luminosity is from a few \% to about 10\% of the
Eddington rate, corresponding to the high/soft state, while the jet is emerged
when the luminosity is nearly equal to the Eddington rate, corresponding to
very high state \citep{Maccarone03,Fender04}.
1H~0323+342 has both highly or mildly relativistic jets in the innermost region
revealed with our observations and extremely high accretion rate with the
accretion disk luminosity of $0.9 L_{\mathrm{Edd}}$ \citep{Abdo09b}.
These imply that the state transition of the source is in accordance with that
of Galactic X-ray binaries.

On the other hand, 1H~0323+342 has very high accretion rate with a smaller
black hole mass of $\sim 10^7 M_\sun$ \citep{Zhou07,Abdo09b}, while it also
shows very radio-loud feature of $R_{1.4} = 318$ \citep{Foschini11}, or
$R_5 = 246$ \citep{Doi12}.
These properties seem to be contrary to previous studies showing a correlation
between radio loudness and the black hole mass
\citep[][and references therein]{McLure04}, and an anti-correlation between
radio loudness and the accretion rate
\citep[][and references therein]{Greene06} for a large sample of AGNs.
Following these correlation, 1H~0323+342 may be intrinsically a radio-quiet one
but seem to be radio-loud as a result of the Doppler-boosting effect, as
similar results have been obtained for several radio-loud NLS1s
\citep{Doi11,Doi12}.
Observed radio flux density is boosted by a factor of
$D = \delta^{3 - \alpha}$, and we obtain $D = 18.5$ applying $\delta = 2.2$ and
$\alpha = -0.70$ derived from our studies.
Thus the intrinsic radio loudness can be estimated as $R_{1.4}^{\mathrm{(int)}}
= 17$ and $R_5^{\mathrm{(int)}} = 13$.
Although these $R$ are still in the radio-loud regime
\citep[$R > 10$;][]{Kellermann89}, 1H~0323+342 intrinsically has much lower $R$
than expected.

\subsection{Radio Spectrum}
\label{subsec:Section5-6}

\begin{figure}
\epsscale{1.00}
\plotone{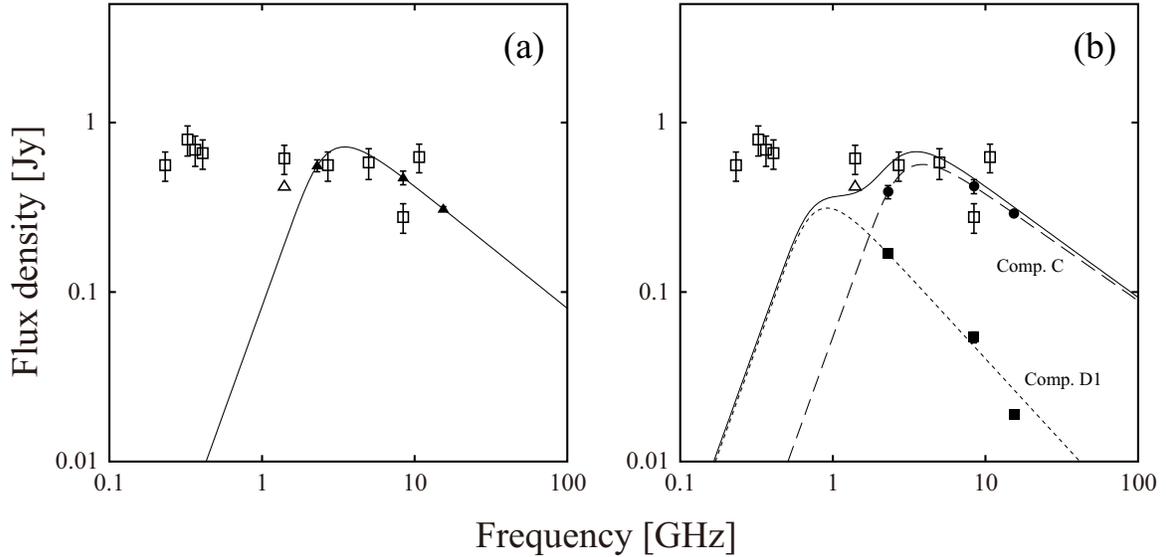}
\caption{
Composite radio spectrum of 1H~0323+342.
The filled symbols represent the VLBI results at 2.3~GHz (VLBA), 8.4~GHz (JVN
epoch 1), and 15.4~GHz (VLBA), while the open squares are the total flux
density by single-dish or radio interferometer.
The open triangle indicates the flux density of the core component obtained by
VLA \citep{Doi12}.
(a) Radio spectrum for the sum of flux for all VLBI components.
The solid line represents the best-fit curve with synchrotron spectrum.
(b) Radio spectra for the components C and D1.
The filled circles and squares are VLBI results for the components C (or
${\rm C0} + {\rm D3} + {\rm D2}$ for VLBA) and D1, respectively.
The dashed and dotted lines represent the best-fit curve with synchrotron
spectrum for the components C and D1, respectively, and the solid line
describes the sum of two synchrotron spectra.
The turnover frequency of 0.9~GHz is assumed for the spectral fitting to the
component D1 (see also Section~\ref{subsec:Section5-7}).
Single-dish and interferometer measurement data are from \citet{Colla73},
\citet{Condon98}, \citet{Douglas96}, \citet{Healey07}, \citet{Neumann94},
\citet{Reich00}, \citet{Rengelink97}, and \citet{Zhang97}.
A part of the flux data is compiled from the SPECFIND V2.0 catalog
\citep{Vollmer10}.
}
\label{fig:Figure7}
\end{figure}

Figure~\ref{fig:Figure7} shows the composite radio spectrum of 1H~0323+342,
including not only our VLBI observation results but also previous single-dish
or interferometric measurements.
We applied spectral fitting to the VLBI data with a synchrotron self-absorption
(SSA) spectral model, $S_{\nu} = S_0 \nu^{2.5} \left[1 - \exp
\left(-\tau_{\mathrm{ss}} \nu^{\alpha - 2.5} \right) \right]$, where $S_0$ is
a scaling constant and and $\tau_{\mathrm{ss}}$ is the SSA coefficient.
In the SSA fit to the sum of the flux of all VLBI components, a peak flux
density, $S_{\mathrm{m}}$, of $0.72$~Jy is derived at a turnover frequency,
$\nu_{\mathrm{m}}$, of 3.5~GHz, as shown in Figure~\ref{fig:Figure7}(a), while
the source is separated into two or more components in
Figures~\ref{fig:Figure2} and \ref{fig:Figure3}.
If we apply the SSA fit to each VLBI component, as shown in
Figure~\ref{fig:Figure7}(b), $S_{\mathrm{m}} = 0.57$~Jy is derived at
$\nu_{\mathrm{m}} = 3.9$~GHz for the component C.
$\nu_{\mathrm{m}}$ can be given under the condition of a homogeneous,
self-absorbed, and incoherent synchrotron radio source with a power-law
electron energy distribution as
\begin{equation}
\nu_{\mathrm{m}} \sim 8 B^{1/5} S_{\mathrm{m}}^{2/5} \theta^{-4/5}
(1+z)^{1/5} \hspace{5mm} {\mathrm{[GHz]}}
\label{eqn:Equation3}
\end{equation}
\citep{Kellermann81}, where $B$~[G] is the magnetic field, and $\theta$~[mas]
is the angular size of a source.
Applying $\nu_{\mathrm{m}} = 3.9$~GHz and $S_{\mathrm{m}} = 0.57$~Jy given by
the SSA fitting results, and $\theta = \sqrt{\theta_{\mathrm{maj}}
\theta_{\mathrm{min}}} \sim 0.4$~mas from VLBI observations, we can derive
$B \sim 2$~mG, which is comparable to the magnetic field for young radio
galaxies \citep[e.g.,][]{Murgia99}.
For the component D1, $S_{\mathrm{m}} = 0.31$~Jy is derived assuming
$\nu_{\mathrm{m}} = 0.9$~GHz in the spectral fitting in
Figure~\ref{fig:Figure7}(b).
Applying $B \sim 2$~mG derived for the component C, and $\theta \sim 2$~mas,
we obtain $S_{\mathrm{m}} \sim 0.37$~Jy from Equation~(\ref{eqn:Equation3}),
which is comparable to that given by the SSA fitting.
Although Equation~(\ref{eqn:Equation3}) depends strongly on the parameters,
the component D1 shows similar $\nu_{\mathrm{m}}$ or $S_{\mathrm{m}}$ between
the spectral fitting and numerical estimation.
In Figure~\ref{fig:Figure7}(b), the sum of two synchrotron spectra accounts for
the total flux measurements at a frequency of greater than 1~GHz.
Especially, the total flux of the central component detected by VLA at 1.4~GHz
\citep[open triangle in Figure~\ref{fig:Figure7};][]{Doi12} is accountable for
the sum of the flux of two VLBI components, implying that the emission of the
VLA central component mainly comes from the pc-scale VLBI components.
However, the composite spectrum for the VLBI components cannot account for the
points at MHz-frequency range.
These emissions probably come from extended, optically-thin components on the
kpc-scale detected with interferometric observations, as mentioned in
Section~\ref{subsec:Section5-7}.

We should point out that the composite radio spectrum shown in
Figure~\ref{fig:Figure7} is not obtained simultaneously.
Future multifrequency, sumultaneous, and high-resolution VLBI observations are
thus important to reveal spectral properties of the innermost region precisely.

\subsection{Possibility of Young Radio Source \\ and Recurrent Jet Activity}
\label{subsec:Section5-7}

Some previous studies point out similarities between NLS1s and young radio
galaxies, such as gigahertz-peaked spectrum sources
\citep[GPSs;][]{O'Dea98}, in terms of compact and stable radio morphologies and
a steep spectrum of an inner region \citep[e.g.,][]{Gallo06,Komossa06}.
As shown in Section~\ref{subsec:Section5-4}, the JVN observations combined with
the archival VLBA data revealed the existence of two stationary or slowly
moving components D2 and D1, and \citet{Zhou07} mention that the component D1
is possibly a weak radio lobe.

The viewing angle of the component, $\phi$, can be obtained using the apparent
velocity, $\beta_{\mathrm{app}} \equiv v_{\mathrm{app}} / c$, and a Doppler
factor, $\delta$, as
\begin{equation}
\phi = \tan^{-1} \frac{2\beta_{\mathrm{app}}}{\beta_{\mathrm{app}}^2
+ \delta^2 -1}
\end{equation}
\citep{Ghisellini93}.
If the component D1 has the same $\delta$ as C and we adopt $\delta = 2.2$, as
derived in Section~\ref{subsec:Section5-2}, we can obtain $\phi = 3\fdg6$,
which is in good agreement with $\phi = 3\degr$  derived by the model fitting
to the spectral energy distribution \citep{Abdo09b}.
We can also calculate the real distance between C and D1 as 0.13~kpc.
Assuming the linear expansion of the components D2 and D1 with the constant
apparent velocity of $0.02 c$ and $0.12 c$, respectively, as shown in
Section~\ref{subsec:Section5-4}, the kinematic age, $t_{\mathrm{kin}}$, of
120 and 220 years are derived for each component.
Similar source age can be estimated from a correlation between the linear size
and $\nu_{\mathrm{m}}$ of a source \citep{O'Dea97}.
Thus we believe that the pc-scale structure in 1H~0323+342 detected by VLBI
observations has GPS-like features in terms of the GHz-peaked spectrum as shown
in Figure~\ref{fig:Figure7}, the young age of $\sim 10^2$ years, and the
compact structure of $\sim 0.1$~kpc.

On the other hand, previous VLA observations show $t_{\mathrm{kin}} \sim 10^7$
-- $10^8$ years for the 100~kpc-scale structure detected by the VLA C-array
configuration \citep{Anton08,Doi12}, while $t_{\mathrm{kin}} \sim 10^6$ --
$10^7$ years can be estimated for the extended structure with the scale of
20~kpc detected by the VLA A-array configuration.
The VLA images also show the extended structure with the position angle of
$\sim 45\degr$ with 100~kpc scale and $\sim 90\degr$ with 20~kpc scale
\citep{Anton08}, both of which are significantly different with that of $\sim
125\degr$ with pc-scale images obtained by VLBI observations.
Additionally, no extended feature could be found by the JVN observations within
the field of view of a few hundred mas (corresponding to a kinematic age of
$\sim 10^4$ years) under the condition of the detection limit as three times
the rms image noise.
These imply that the pc-scale jet structure represents recurrent jet activity
\citep[e.g.,][]{Baum90,Augusto06,Doi13b}.
In fact, the 8~GHz radio power for the pc-scale components is
$P_{\mathrm{8GHz}} = 10^{24.6}$~W~Hz$^{-1}$, as shown in
Section~\ref{sec:Section4}, and the jet kinetic power is roughly estimated to
be $\sim 10^{44}$~erg~s$^{-1}$ from $P_{\mathrm{8GHz}}$, which is enough to
make the source grow to the size of a hundred kpc \citep{Doi12,Doi13a}.
Extremely large difference in the position angle between kpc- and pc-scales may
be the projection effect of extended components due to the small viewing angle.

\subsection{Possibility of Free-Free Absorption}
\label{subsec:Section5-8}

Free-free absorption could be an explanation to make an apparent asymmetric
structure, and previous VLBI observations have revealed the existence of
free-free absorption in several GPSs \citep[e.g.,][]{Kameno00,Kameno01,Marr01}.
The optical depth of free-free absorption can be estimated as
\begin{equation}
\tau_{\mathrm{ff}} = 8.235 \times 10^{-2} T_{\mathrm{e}}^{-1.35}
\nu^{-2.1} \int_{\mathrm{LOS}} n_{\mathrm{e}}^2 dl
\end{equation}
\citep{Mezger67}, where $T_{\mathrm{e}}$~[K] and $n_{\mathrm{e}}$~[cm$^{-3}$]
are the electron temperature and density of the absorber, $\nu$~[GHz] is the
frequency, and $l$~[pc] is the path length via the line of sight.
Peak intensity of the component D1 is 38~mJy~beam$^{-1}$ on the image of JVN
epoch 1, resulting in the attenuation of more than a factor of 4.1
if we assume the intrinsic jet symmetry and the detection limit of the counter
component as three times the rms image noise.
The $\tau_{\mathrm{ff}}$ of greater than 1.4 is thus required to reconcile with
the observed asymmetry.
Assuming $T_{\mathrm{e}} > 8000$~K for the fully ionized condition,
$l = 0.13$~kpc for the distance between the components C and D1, and
$\nu = (1+z) \times 8.4 = 8.9$~GHz, the column density of electrons
$N_{\mathrm{e}} = n_{\mathrm{e}} l > 6 \times 10^{23}$~cm$^{-2}$ is required.
This $N_{\mathrm{e}}$ is two orders of magnitude larger than the Galactic
\ion{H}{1} column density of $1.349 \times 10^{21}$~cm$^{-2}$ obtained by the
{\it ROSAT} All-Sky Survey \citep{Voges99}.
Moreover, the kpc-scale radio morphology of the source seems to be an
asymmetric structure elongated to the east, as shown by \citet{Anton08},
although they consider the morphology to be a core plus a two-sided structure.
As shown in Section~\ref{subsec:Section5-5}, we obtain $D = 18.5$ as a boosting
factor for the radio flux.
This $D$ is accountable for the observed flux ratio of approaching to receding
jet components mentioned above.
We therefore conclude that the asymmetric structure detected by the JVN
observations is not due to free-free absorption, but due to the Doppler-beaming
effect.

\subsection{Gamma-Ray Emission from 1H~0323+342}
\label{subsec:Section5-9}

\citet{Marscher08} analyzed the multiwavelength light curve and
quasi-simultaneous multi-epoch VLBI images of a blazar BL Lacertae, and
proposed the inner jet model to explain the flare timing between each energy
band and sudden change of the optical polarization position angle.
They suggest that the flare is stimulated by the passage of the emission
feature through the conical standing shock at around $10^5 R_{\mathrm{s}}$ from
the central black hole, where $R_{\mathrm{s}}$ is the Schwarzschild radius, and
the gamma-ray flare would occur at this area.
It corresponds to a linear scale of 0.1~pc or an angular size of 0.08~mas for
1H~0323+342.
Although our observations combined with the archival VLBA results reveal the
existence of the relativistic jet in the innermost region and stationary or
slowly moving components D2 and D1, as shown in
Section~\ref{subsec:Section5-4},
the linear distance of D2 and D1 from C0 is $10^7 R_{\mathrm{s}}$ and $10^8
R_{\mathrm{s}}$, respectively, assuming $\phi$ from our results.
The central black hole may be located upstream of the core at 8~GHz as a result
of the frequency-dependent position shift, thus the distance between the
central black hole and the components D2 and D1 is farther than those mentioned
above.
On the other hand, the linear scale of $10^5 R_{\mathrm{s}}$ is comparable to
the light-crossing distance corresponding to the time scale of the short-term
radio variability ($0.3 \times 10^5 R_{\mathrm{s}}$), which is probably
associated with the component C (and the gamma-ray emitting region), as shown
in Section~\ref{subsec:Section3-2}.
Moreover, the VLBA image at 15~GHz shown in Figure~\ref{fig:Figure3} indicates
the existence of a component D3 in the vicinity of the component C0 with a
separation angle of 0.19~mas, or a linear distance of $2 \times 10^5
R_{\mathrm{s}}$.
Proper motion and flux variation of the component D3 might be related to the
gamma-ray flux, and future high-resolution and high-sensitivity VLBI
observations will be important to investigate the gamma-ray emission mechanism
from NLS1s.

Gamma-ray emission has been detected from blazars with larger Doppler factor
and smaller viewing angle \citep[e.g., $\delta \ga 14$ and $\phi < 4\fdg8$ for
PKS~1741$-$038;][]{Wajima00}, and with smaller Doppler factor and larger
viewing angle \citep[e.g., $\delta \sim 2.5$ and $\phi = 23\fdg3$ for
PKS~1622$-$297;][]{Wajima06}.
Both blazars have superluminal jet components and thus show very high intrinsic
velocities with $\beta > 0.9$, implying that the inverse-Compton process plays
a key role for gamma-ray emission.
On the other hand, 1H~0323+342 has smaller $\delta$, $\beta$, and $\phi$ and
thus seems to have different features compared to typical gamma-ray blazars.
Future simultaneous, multiwavelength observations from radio to gamma-ray and
high-resolution polarization studies will be key issues to reveal the gamma-ray
emission mechanism from NLS1s.

\section{Conclusion}
\label{sec:Section6}

We made simultaneous single-dish and VLBI observations of a gamma-ray NLS1
galaxy 1H~0323+342 at 8~GHz.
The achievements of our study can be summarized as follows:
\begin{enumerate}
\item We found significant flux variation on the time scale of one month with
the single-dish monitoring by Yamaguchi 32~m radio telescope.
The total flux density varied by 5.5\% in 32~days, corresponding to a
variability brightness temperature of $7.0 \times 10^{11}$~K.
\item Milliarcsecond-scale images obtained by three-epoch Japanese VLBI Network
observations show that the source has a compact core-jet structure similar to
that of blazars.
The visibilities can be modeled satisfactorily by two elliptical Gaussian
components.
Only the central component shows flux decrease similar to that of the total
flux obtained with the single-dish monitoring, while the flux density of the
southeastern component seems to be stable.
\item Two-year {\it Fermi}/LAT monitoring results show that the source has
gamma-ray flux variation on the similar time scale to our single-dish
monitoring results.
By combining the results 1. and 2., we conclude that the source of short-term
radio variability is probably associated with the gamma-ray emitting region.
\item The brightness temperature obtained by the JVN observations is greater
than $(5.2 \pm 0.3)\times 10^{10}$~K, and the radio power at 8~GHz is estimated
to be $10^{24.6}$~W~Hz$^{-1}$.
These indicate that a nonthermal process in the central region is responsible
for the radio emission from the source.
\item The Doppler factor is estimated to be $\delta_{\mathrm{eq}} > 1.7$
assuming the condition of energy equipartition in the JVN observations, and
$\delta_{\mathrm{var}} = 2.2$ from the observed radio variability, indicating
the existence of relativistic jet(s).
This is the third source in which the Doppler beaming effect is detected in
gamma-ray NLS1s by both direct imaging with VLBI and the flux variation.
\item The JVN observations revealed the existence of Doppler-boosted jet
components which affect radio loudness of the source.
Although the source shows extremely high apparent radio loudness with $R >
200$, the intrinsic $R$ is less than 20 applying the Doppler factor obtained by
the JVN observations.
Hence, for 1H~0323+342 with a smaller black hole mass ($\sim 10^7 M_{\sun}$)
and high accretion rate ($\sim 0.9 L_{\mathrm{Edd}}$), there seems to be no
large discrepancy with previous studies suggesting a correlation between $R$
and the black hole mass, and anti-correlation between $R$ and the accretion
rate.
\item Multiepoch JVN and VLBA images detected slowly moving jet components D2
and D1 with apparent velocities of $(0.02 \pm 0.03)c$ and $(0.12 \pm 0.04)c$,
respectively, and the real distance between the components C and D1 was
estimated as 0.13~kpc.
The kinematic age of $\sim 10^2$ years is derived for both components assuming
the linear expansion, which is significantly different with that of kpc-scale
components ($\sim 10^7$ -- $10^8$ years).
By combining the results of the spectral fitting to the flux measurements and
observed properties in terms of the position angle and lack of medium-scale
radio emission, we conclude that the pc-scale jet structure represents
recurrent jet activity.
\end{enumerate}

\acknowledgments

We are grateful to the anonymous referee for valuable comments which improved
the manuscript.
The JVN project is led by the National Astronomical Observatory of Japan
(NAOJ), which is a branch of the National Institutes of Natural Sciences
(NINS), Hokkaido University, Ibaraki University, University of Tsukuba, Gifu
University, Osaka Prefecture University, Yamaguchi University, and Kagoshima
University, in cooperation with the Geographical Survey Institute (GSI), the
Japan Aerospace Exploration Agency (JAXA), and the National Institute of
Information and Communications Technology (NICT).
The VLBA is operated by the National Radio Astronomy Observatory, which is a
facility of the National Science Foundation operated under cooperative
agreement by Associated Universities, Inc.
This research has made use of the following data, tools, and facilities;
the data from the MOJAVE database that is maintained by the MOJAVE team
\citep{Lister09}, the Swinburne University of Technology software correlator
\citep{Deller11}, NASA's Astrophysics Data System Abstract Service, the
NASA/IPAC Extragalactic Database (NED), which is operated by the Jet Propulsion
Laboratory, Ned Wright's on-line cosmology calculator, and the VizieR catalogue
access tool, CDS, Strasbourg, France \citep{Ochsenbein00}.
This work is partly supported by the National Natural Science Foundation of
China (grant 11121062), the CAS/SAFEA International Partnership Program for
Creative Research Teams, and the Strategic Priority Research Program on Space
Science, the Chinese Academy of Sciences (Grant No.\ XDA04060700).


\clearpage

\clearpage

\begin{deluxetable}{cllcccccc}
\tabletypesize{}
\tablewidth{0pt}
\tablecaption{Summary of JVN observations and description of JVN images.\label{tbl:Table1}}
\tablehead{
\colhead{Epoch} & \colhead{Date (UT)} & \colhead{Telescopes} & \colhead{$\theta_{\mathrm{Bmaj}}$} & \colhead{$\theta_{\mathrm{Bmin}}$} & \colhead{PA}    & \colhead{$S_{\mathrm{peak}}$}    & \colhead{$\sigma$}          & \colhead{$S_{\mathrm{CLEAN}}$} \\
\colhead{}      &                     &                      & \colhead{[mas]}                    & \colhead{[mas]}                    & \colhead{[deg]} & \colhead{[mJy~beam$^{-1}$]}      & \colhead{[mJy~beam$^{-1}$]} & \colhead{[mJy]} \\
\colhead{(1)}   & \colhead{(2)}       & \colhead{(3)}        & \colhead{(4)}                      & \colhead{(5)}                      & \colhead{(6)}   & \colhead{(7)}                    & \colhead{(8)}               & \colhead{(9)}}
\startdata
1 & 2010 Nov \phn1 11:00 -- 21:00 & VERA, HIT      & 3.52 & 1.81 & 124.0 & 397 & 3.1 & 467 \\
2 & 2010 Nov 15    10:00 -- 20:00 & VERA, HIT, KAS & 3.73 & 1.85 & 126.0 & 298 & 2.8 & 359 \\
3 & 2010 Nov 30    09:00 -- 19:00 & VERA, HIT, KAS & 3.77 & 1.80 & 124.5 & 272 & 2.2 & 335
\enddata
\tablecomments{Column 1: JVN observation epoch; Column 2: observation date; Column 3: telescopes: VERA -- VERA (4 $\times$ 20~m), HIT -- Hitachi 32~m,
KAS -- Kashima 34~m; Columns 4, 5, and 6: parameters of the restoring beam: Full-width at half maximum (FWHM) of major and minor axes and the position
angle of the major axis; Column 7: peak intensity; Column 8: rms noise level; Column 9: total CLEANed flux.}
\end{deluxetable}

\begin{deluxetable}{lrrr}
\tabletypesize{}
\tablewidth{0pt}
\tablecaption{Results of the total flux measurement at 8.38~GHz
with Yamaguchi 32~m radio telescope.\label{tbl:Table2}}
\tablehead{
\colhead{Date} & \colhead{Epoch} & \colhead{$S_{\mathrm{8GHz}}$} & \colhead{$N$\tablenotemark{a}} \\
               &                 & \colhead{[mJy]}               & }
\startdata
2010 Nov  9 & 2010.858 & $427 \pm    28$ & 31 \\
2010 Nov 13 & 2010.868 & $432 \pm    24$ & 35 \\
2010 Nov 17 & 2010.879 & $366 \pm    27$ & 32 \\
2010 Nov 21 & 2010.890 & $359 \pm    16$ & 39 \\
2010 Nov 23 & 2010.896 & $369 \pm    15$ & 38 \\
2010 Dec  1 & 2010.918 & $358 \pm    18$ & 39 \\
2010 Dec  3 & 2010.923 & $374 \pm    11$ & 39 \\
2010 Dec  5 & 2010.929 & $325 \pm    12$ & 39 \\
2010 Dec  7 & 2010.934 & $379 \pm    27$ & 38 \\
2010 Dec  9 & 2010.940 & $288 \pm    18$ & 38 \\
2010 Dec 11 & 2010.945 & $309 \pm    13$ & 40 \\
2010 Dec 18 & 2010.964 & $305 \pm \phn8$ & 39 \\
2010 Dec 19 & 2010.967 & $345 \pm    26$ & 26 \\
2010 Dec 23 & 2010.978 & $324 \pm \phn9$ & 36 \\
2010 Dec 26 & 2010.986 & $373 \pm    25$ & 24 \\
2010 Dec 29 & 2010.995 & $355 \pm    21$ & 31 \\
2010 Dec 30 & 2010.997 & $340 \pm    13$ & 39 \\
2011 Jan  4 & 2011.011 & $350 \pm    21$ & 26 \\
2011 Jan  6 & 2011.016 & $363 \pm    17$ & 36 \\
2011 Jan  8 & 2011.022 & $287 \pm    24$ & 25 \\
2011 Jan 18 & 2011.049 & $348 \pm    21$ & 17 \\
2011 Jan 24 & 2011.066 & $363 \pm    23$ & 18 \\
2011 Jan 26 & 2011.071 & $338 \pm    35$ & 12 \\
2011 Feb  3 & 2011.093 & $324 \pm    20$ & 15 \\
2011 Feb  5 & 2011.099 & $266 \pm    30$ & 12
\enddata
\tablenotetext{a}{Number of independent measurements.}
\end{deluxetable}

\begin{deluxetable}{ccrccrrrc}
\tabletypesize{}
\tablewidth{0pt}
\tablecaption{Model fitting results for JVN observations.\label{tbl:Table3}}
\tablehead{
\colhead{Epoch} & \colhead{Comp.} & \colhead{$S$}   & \colhead{$r$}   & \colhead{$\phi$} & \colhead{$\theta_{\mathrm{maj}}$} & \colhead{$\theta_{\mathrm{min}}$} & \colhead{PA}    & \colhead{$T_{\mathrm{B,rest}}^{\mathrm{(image)}}$} \\
\colhead{}      & \colhead{}      & \colhead{[mJy]} & \colhead{[mas]} & \colhead{[deg]}  & \colhead{[mas]}                   & \colhead{[mas]}                   & \colhead{[deg]} & \colhead{[$10^{10}$~K]} \\
\colhead{(1)}   & \colhead{(2)}   & \colhead{(3)}   & \colhead{(4)}   & \colhead{(5)}    & \colhead{(6)}                     & \colhead{(7)}                     & \colhead{(8)}   & \colhead{(9)}}
\startdata
1 & C  & $   419 \pm    40$ & \nodata         & \nodata         & $ < 1.12 \pm 0.11$ & $ < 0.12 \pm 0.01$ & $   131.1 \pm 0.4$ & $> 8.3 \pm 1.3$ \\
  & D1 & $\phn54 \pm \phn4$ & $7.22 \pm 0.17$ & $122.8 \pm 2.1$ & $   3.76 \pm 0.38$ & $   0.11 \pm 0.01$ & $   116.5 \pm 2.7$ & \nodata         \\
2 & C  & $   309 \pm    12$ & \nodata         & \nodata         & $ < 0.59 \pm 0.02$ & $ < 0.24 \pm 0.01$ & $\phn84.6 \pm 0.5$ & $> 5.8 \pm 0.4$ \\
  & D1 & $\phn47 \pm \phn1$ & $6.25 \pm 0.13$ & $125.8 \pm 1.9$ & $   2.76 \pm 0.11$ & $   0.27 \pm 0.01$ & $   109.6 \pm 3.4$ & \nodata         \\
3 & C  & $   279 \pm    11$ & \nodata         & \nodata         & $ < 0.55 \pm 0.02$ & $ < 0.26 \pm 0.01$ & $   117.9 \pm 0.4$ & $> 5.2 \pm 0.3$ \\
  & D1 & $\phn49 \pm \phn1$ & $6.26 \pm 0.14$ & $124.2 \pm 1.4$ & $   5.22 \pm 0.21$ & $   1.62 \pm 0.07$ & $   134.4 \pm 3.6$ & \nodata
\enddata
\tablecomments{Column 1: JVN observation epoch; Column 2: component name; Column 3: flux density; Column 4: distance from the origin defined by component C;
Column 5: position angle with respect to the origin (measured from north through east); Columns 6, 7, and 8: parameters of Gaussian model: FWHM of major and
minor axes and the position angle of the major axis; Column 9: brightness temperature given in the source's rest frame.}
\end{deluxetable}

\begin{deluxetable}{lllcrccrccc}
\tabletypesize{}
\tablewidth{0pt}
\tablecaption{Observation summary and map description of VLBA images.\label{tbl:Table4}}
\tablehead{
\colhead{Code} & \multicolumn{2}{c}{Date (epoch)} & \colhead{$\nu_{\mathrm{obs}}$} & \colhead{$t_{\mathrm{on}}$} & \colhead{$\theta_{\mathrm{Bmaj}}$} & \colhead{$\theta_{\mathrm{Bmin}}$} & \colhead{PA}    & \colhead{$S_{\mathrm{peak}}$} & \colhead{$\sigma$}          & \colhead{Contours} \\
\colhead{}     & \multicolumn{2}{c}{}             & \colhead{[GHz]}                & \colhead{[sec]}             & \colhead{[mas]}                    & \colhead{[mas]}                    & \colhead{[deg]} & \colhead{[mJy~beam$^{-1}$]}   & \colhead{[mJy~beam$^{-1}$]} & \colhead{} \\
\colhead{(1)}  & \multicolumn{2}{c}{(2)}          & \colhead{(3)}                  & \colhead{(4)}               & \colhead{(5)}                      & \colhead{(6)}                      & \colhead{(7)}   & \colhead{(8)}                 & \colhead{(9)}               & \colhead{(10)}}
\startdata
BB023 & 1996 May 16 & (1996.374) & \phn2.269 &  322 & 6.00 & 2.77 &    9.0  & 351 & 2.6 & $n=0$, 1, $\cdot\cdot\cdot$, 11 \\
      & 1996 May 16 & (1996.374) & \phn8.339 &  322 & 1.73 & 0.73 &    8.7  & 245 & 1.0 & $n=0$, 1, $\cdot\cdot\cdot$, 12 \\
BK077 & 2003 Oct  9 & (2003.688) & \phn8.421 &  178 & 2.26 & 1.35 & $-16.9$ & 334 & 2.5 & $n=0$, 1, $\cdot\cdot\cdot$, 14 \\
BE042 & 2005 Aug 20 & (2005.636) & \phn8.420 & 6640 & 1.93 & 0.63 & $ -2.0$ & 204 & 1.3 & $n=0$, 1, $\cdot\cdot\cdot$, 14 \\
BL156 & 2008 Apr 20 & (2008.303) & \phn8.392 &  281 & 1.98 & 0.81 & $ -7.5$ & 217 & 1.6 & $n=0$, 1, $\cdot\cdot\cdot$, 14 \\
BL149 & 2010 Oct 15 & (2010.789) &    15.357 & 2088 & 0.76 & 0.47 &    0.0  & 223 & 0.3 & $n=0$, 1, $\cdot\cdot\cdot$, 15
\enddata
\tablecomments{Column 1: VLBA observation code; Column 2: observation date and epoch; Column 3: observation frequency; Column 4: total on-source time;
Columns 5, 6, and 7: parameters of the restoring beam: Full-width at half maximum (FWHM) of major and minor axes and the position angle of the major axis;
Column 8: peak intensity; Column 9: rms noise level; Column 10: contour levels. The number $n$ corresponds to $-3\sigma$, $3\sigma \times (\sqrt{2})^n$.}
\end{deluxetable}

\begin{deluxetable}{crccccccc}
\tabletypesize{}
\tablewidth{0pt}
\tablecaption{Model fitting results for archival VLBA observations.\label{tbl:Table5}}
\tablehead{
\colhead{Epoch} & \colhead{$\nu$} & \colhead{Comp.} & \colhead{$S$}   & \colhead{$r$}   & \colhead{$\phi$} & \colhead{$\theta_{\mathrm{maj}}$} & \colhead{$\theta_{\mathrm{min}}$} & \colhead{PA} \\
\colhead{}      & \colhead{[GHz]} & \colhead{}      & \colhead{[mJy]} & \colhead{[mas]} & \colhead{[deg]}  & \colhead{[mas]}                   & \colhead{[mas]}                   & \colhead{[deg]} \\
\colhead{(1)}   & \colhead{(2)}   & \colhead{(3)}   & \colhead{(4)}   & \colhead{(5)}   & \colhead{(6)}    & \colhead{(7)}                     & \colhead{(8)}                     & \colhead{(9)}}
\startdata
1996.374 & 2.269  & C0+D2 & $   389 \pm    35$ & \nodata         & \nodata         & $1.33 \pm 0.13$ & $1.00 \pm 0.10$ & $\phn55.9 \pm \phn0.4$ \\
         &        & D1    & $   168 \pm    10$ & $6.39 \pm 1.31$ & $126.7 \pm 1.2$ & $3.80 \pm 0.38$ & $2.22 \pm 0.22$ & $   147.4 \pm \phn1.7$ \\
1996.374 & 8.339  & C0    & $   271 \pm    11$ & \nodata         & \nodata         & $0.49 \pm 0.02$ & $0.14 \pm 0.01$ & $   125.9 \pm \phn0.2$ \\
         &        & D2    & $\phn51 \pm \phn1$ & $0.58 \pm 0.26$ & $110.6 \pm 0.4$ & $1.37 \pm 0.07$ & $0.13 \pm 0.01$ & $   131.8 \pm \phn3.0$ \\
         &        & D1    & $\phn40 \pm \phn1$ & $6.70 \pm 0.32$ & $125.0 \pm 2.4$ & $3.27 \pm 0.16$ & $1.43 \pm 0.07$ & $   138.6 \pm \phn8.2$ \\
2003.688 & 8.421  & C0+D2 & $   348 \pm    13$ & \nodata         & \nodata         & $0.52 \pm 0.02$ & $0.35 \pm 0.01$ & $   120.5 \pm \phn0.2$ \\
         &        & D1    & $\phn40 \pm \phn1$ & $7.03 \pm 0.31$ & $126.3 \pm 5.4$ & $3.36 \pm 0.13$ & $0.62 \pm 0.03$ & $   121.0 \pm \phn2.4$ \\
2005.636 & 8.420  & C0    & $   203 \pm \phn6$ & \nodata         & \nodata         & $0.35 \pm 0.01$ & $0.15 \pm 0.01$ & $   148.4 \pm \phn0.1$ \\
         &        & D2    & $\phn43 \pm \phn1$ & $0.50 \pm 0.21$ & $124.6 \pm 0.2$ & $1.48 \pm 0.04$ & $0.15 \pm 0.01$ & $   117.8 \pm \phn1.7$ \\
         &        & D1b   & $\phn13 \pm \phn1$ & $6.93 \pm 0.20$ & $124.8 \pm 2.0$ & $0.93 \pm 0.02$ & $0.30 \pm 0.01$ & $   107.1 \pm \phn1.5$ \\
         &        & D1a   & $\phn31 \pm \phn1$ & $7.18 \pm 0.21$ & $125.4 \pm 2.3$ & $5.86 \pm 0.33$ & $1.17 \pm 0.01$ & $   127.3 \pm \phn2.0$ \\
2008.303 & 8.392  & C0    & $   179 \pm \phn5$ & \nodata         & \nodata         & $0.30 \pm 0.01$ & $0.09 \pm 0.01$ & $   137.0 \pm \phn0.1$ \\
         &        & D2    & $\phn98 \pm \phn2$ & $0.57 \pm 0.21$ & $130.7 \pm 0.2$ & $1.35 \pm 0.04$ & $0.22 \pm 0.01$ & $   117.6 \pm \phn0.6$ \\
         &        & D1b   & $\phn29 \pm \phn1$ & $6.86 \pm 0.28$ & $126.3 \pm 5.1$ & $5.11 \pm 0.15$ & $1.09 \pm 0.03$ & $   127.0 \pm \phn3.1$ \\
         &        & D1a   & $\phn12 \pm \phn1$ & $7.30 \pm 0.21$ & $126.0 \pm 2.4$ & $1.56 \pm 0.05$ & $0.44 \pm 0.01$ & $   137.3 \pm \phn2.6$ \\
2010.789 & 15.357 & C0    & $   193 \pm \phn4$ & \nodata         & \nodata         & $0.18 \pm 0.01$ & $0.04 \pm 0.01$ & $   148.5 \pm \phn0.1$ \\
         &        & D3    & $\phn48 \pm \phn4$ & $0.19 \pm 0.19$ & $119.4 \pm 0.1$ & $0.20 \pm 0.01$ & $0.09 \pm 0.01$ & $   142.7 \pm \phn0.9$ \\
         &        & D2    & $\phn49 \pm \phn1$ & $0.67 \pm 0.24$ & $126.3 \pm 0.2$ & $1.97 \pm 0.08$ & $0.29 \pm 0.01$ & $   131.5 \pm \phn2.1$ \\
         &        & D1    & $\phn19 \pm \phn1$ & $7.10 \pm 0.27$ & $125.0 \pm 2.4$ & $2.77 \pm 0.11$ & $0.67 \pm 0.03$ & $   124.4 \pm    11.8$
\enddata
\tablecomments{Column 1: VLBA observation epoch; Column 2: observation frequency; Column 3: component name; Column 4: flux density;
Column 5: distance from the origin defined by component C0; Column 6: position angle with respect to the origin (measured from north through east);
Columns 7, 8, and 9: parameters of Gaussian model: FWHM of major and minor axes and the position angle of the major axis.}
\end{deluxetable}

\end{document}